\documentclass[12pt]{iopart}
\usepackage{graphicx}
\usepackage{subcaption}
\begin{document}

	\title[]{Numerical simulations of electron acceleration driven by heavy ion beams in plasma with alternating density gradients}
	
	\author{
		Jiangdong Li$^{1,2}$, 
		Jiancheng Yang$^{1,2}$\footnote{Present address: Institute of Modern physics, Chinese Academy of Sciences, Lanzhou 730000, China},
		Guoxing Xia$^{3,4}$,
		Jie Liu$^{1,2}$,
		Ruihu Zhu$^{1,2}$
		Xiangwen Qiao$^{1,2}$
	}
	
	\address{
		$^1$ Institute of Modern physics, Chinese Academy of Sciences, Lanzhou 730000, China \\
		$^2$ University of Chinese Academy of Sciences, Beijing 100049, China \\
		$^3$ University of Manchester, Manchester M13 9PL, United Kingdom \\
		$^4$ Cockcroft Institute, Daresbury, Cheshire WA4 4AD, United Kingdom
	}
	\ead{yangjch@impcas.ac.cn}
	
	\vspace{10pt}
	\begin{indented}
		\item[]June 2025
	\end{indented}
	
	\begin{abstract}
		Plasma-Based Acceleration (PBA) has been demonstrated using laser, electron, and proton drivers. However, significant challenges remain in achieving high efficiency, stable acceleration, and scalable energy gain. Heavy ion beam drivers, with their high kinetic energy, offer the potential for greater energy transfer to the witness beam. Unfortunately, limited by the relatively low velocity of heavy ion, the dephasing length is really short leading to a low energy gain of the witness beam. Conventional method that increase plasma density linearly is ineffective in this context because the mismatch between the RMS beam radius and plasma wavelength will make the wakefield degrade or even disappear. In this paper, we propose a method that periodically switches the witness beam between different accelerating phase, allowing it to shift between adjacent accelerating cavities. Therefore, the plasma density does not only strictly increase, but also decrease. This will help maintain the structure of wakefield and increase the energy gain of the witness beam.
	\end{abstract}
	
	%\maketitle
	
	\section{Introduction}
	
	In high energy physics, the primary objective is to explore and discover the fundamental laws of nature, including the basic constituents of matter and the interactions between them. To achieve this goal, high energy particle accelerators serve as indispensable tools. As our exploration of nature deepens, there is a growing demand for more advanced and powerful accelerators capable of probing ever more precise and higher energy regimes. Currently, the Large Hadron Collider (LHC) at European Organization for Nuclear Research (CERN) is the highest energy accelerator in the world. It can provide proton-proton collisions at 13 TeV center-of-mass energy within a 27 km ring. Building on this remarkable achievement, the Future Circular Collider (FCC) \cite{FCC:2018byv}, CERN's ambitious next generation accelerator, is designed to far exceed the capabilities of the LHC up to an unprecedented 100 TeV. This would help scientists probe new physics frontiers including dark matter research and detailed investigations into the fundamental properties of the Higgs boson. However, this groundbreaking project comes with significant challenges in both scale and resources. With a planned of 91 km of circumference, FCC is expected to require tens of billions euros. Moreover, the timeline for such projects are long. LHC took more than 30 years from concept to realization and FCC would take 70 years on R\&D and the construction of two stages.
	
	To keep future higher energy colliders in an affordable size and cost, researchers are actively exploring a frontier technique named plasma-based acceleration (PBA). Proposed first by Tajima and Dawson \cite{Tajima:1979bn}, PBA use intense laser or particle beams to generate ultra high electric fields within plasma. Unlike traditional particle accelerators, which rely on kilometer-scale radiofrequency (RF) cavities to gradually boost particle energies, these plasma wakefield can sustain acceleration gradients 10000 times higher (up to $\sim$ TV/m) than conventional RF-based methods. PBA could potentially shorten future colliders from tens of kilometers to just a few hundred meters while reaching comparable or even higher energies.
	
	Due to the availability of ultrashort and high intensity laser pulses, laser wakefield acceleration (LPWA) was proposed and explored first \cite{Esarey:2009zz,Pukhov:2002otp,Lu:2006nz}. In recent experiments at BELLA Center in Berkeley Lab, using high quality guiding of 500 TW laser pulses, electron bunches can be accelerated up to 10 GeV over 30 cm in plasma \cite{Picksley:2024cdd}. Although LPWA can achieve ultra-high wakefield amplitude, the energy gain is relatively low in one single stage due to defocusing and energy depletion of laser pulses. In order to accelerate an electron bunch to TeV, these acceleration gradients would have to be maintained over tens of meters, or many acceleration stages would have to be combined. There are some great challenges for relative timing, alignment, and matching between these acceleration stages. Besides that, the effective acceleration gradient is reduced due to the introduction of long sections, which are used for injections of the driver beam and dumping of the spent beam between the acceleration stages.
	
	Later, it was recognized that relativistic charged particle beam, such as electron beams, can also excited high amplitude plasma wakefields \cite{Chen:1984up,Blue:2003nk}. The first experiment of beam driven plasma wakefield acceleration (PWFA) was conducted at SLAC \cite{Blumenfeld:2007ph}. This experiment demonstrated an electron beam can excite a wakefield of 52 GV/m within 85 cm-long plasma. Then, a follow-up experiment at SLAC successfully accelerated a trailing electron bunch to 1.6 GeV over 36 cm, achieving a good energy spread of 0.7 \% \cite{Litos:2014yqa}. In this experiment the energy transfer efficiency from the wake to electron bunch exceeded 30 \%. However, the transformer ratio, which is less than 2, constrained the maximum energy gain for the witness beam in any symmetric bunch distribution \cite{Ruth:1984pz}. Meanwhile, AWAKE (Advanced Wakefield Experiment) at CERN proposed and then successfully demonstrated proton-driven plasma wakefield acceleration \cite{Muggli:2016pou,Adli:2016rwp,Caldwell:2015rkk}. This experiment successfully accelerated electrons to 2 GeV in 10 m-long plasma. Owing to the high kinetic energy of proton bunch, this technique holds the potential to accelerate witness electrons to the TeV scale in one single acceleration stage \cite{AWAKE:2018gdq}.
	
	In our previous work \cite{li2025numericalinvestigationsheavyion}, it can be seen that heavy ion beam presents several potential advantages in plasma wakefield acceleration. High beam charge density means heavy ion beam can excite wakefields with higher amplitude. Additionally, with larger particle mass, they can sustain stable wakefields structure over longer distances. And compared to laser pulse, electron and proton beam, the most distinguishing characteristic of heavy ion beam is its high kinetic energy. If this energy could fully transfer to the accelerated beam, it could achieve an extremely high energy gains. Unfortunately, limited by the velocity of heavy ion beam, the dephasing length of the accelerated electron beam is really short and the energy gain for one acceleration stage is relatively low. Thus, a critical challenge for heavy ion driven plasma wakefield acceleration is to extend the dephasing length.
	
	This paper is organized as follows: Section 2 analyzes the establishment of plasma wakefields. Section 3 introduces the proposed design method. Section 4 presents simulation results for electron acceleration under various conditions. Conclusions are summarized in Section 5.
	
	\section{Principle of plasma wakefield excitation}
	
\begin{figure}[h]
		\centering
		\includegraphics[width=8.6cm]{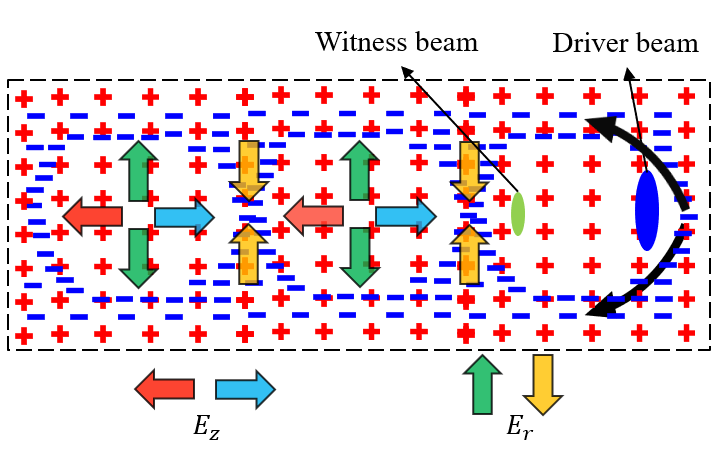}
		\caption{\label{fig:Figure_1}(Color) Illustration of the basic principles of a plasma wakefield accelerator.}
	\end{figure}

	The basic principle of a plasma accelerator is illustrated in figure \ref{fig:Figure_1}. When a driver beam propagates through a plasma, for laser pulses and negative charged particle beams like electron, the pondermotive force from laser pulses and the space charge fields from electron beams expel the plasma electrons from the beam path. If the beam charge density is higher than the plasma density, nearly all plasma electrons will be expelled, which is known as the blowout regime. Due to the heavier mass of plasma background ions, they will not move significantly, maintaining a quasi-stationary state. After the leave of the driver beam, the ions attract the expelled plasma electrons back toward the propagation axis, creating a plsama electron density oscillation, commonly referred to as a plasma wake. The phase velocity of the wake is equal to the velocity of the driver beam. In this wake the plasma background ions contain strong focusing fields and the plasma electrons which are expelled out creates very strong accelerating fields. In an ideal plasma accelerator, one accelerated beam (called the witness beam) will be injected to a appropriate phase. This allows the wakefields to efficiently transfer energy from the driver beam to the witness beam, enabling rapid acceleration and good beam quality.
	
	According to \cite{AWAKE:2023pqh}, when the RMS beam radius of the driver beam $\sigma_r$ is longer than the plasma skin depth $\delta = c / \omega_{pe}$, the plasma return current will flow within the driver beam. Here, $\omega_{pe} = \sqrt{{n_{pe}e^2}\left(\epsilon_0 m_e\right)^{-1}}$ is the plasma electron oscillation frequency, related to the plasma electron density $n_{pe}$. This leads to the development of the current filamentation instability (CFI). CFI will modulate the driver beam transversely into multiple filament and suppress the development of plasma wakefield.
	
	Therefore, to drive plsama wakefield efficiently, the RMS beam radius of the driver beam should satisfy
	
	\begin{equation}\label{eq:satisfy_condition1}
		\delta = \frac{c}{\omega_{pe}} \ge \sigma_r
	\end{equation}
	or
	
	\begin{equation}\label{eq:satisfy_condition2}
		\frac{2\pi \sigma_r}{\lambda_{pe}} = k_{pe} \sigma_r \le 1.
	\end{equation}
	where $\lambda_{pe}$ is the plasma electron wavelength. The achievable accelerating field, called the wave breaking field, $E_{WB}$:
	
	\begin{equation}\label{eq:wave_breaking_field1}
		E_{WB} = \frac{m_e c \omega_{pe}} {e} = \sqrt{\frac{n_{pe} m_e c^2}{\epsilon_0}}
	\end{equation}
	or, the wave breaking field $E_{WB}$ can be approximated as
	
	\begin{equation}\label{eq:wave_breaking_field2}
		E_{WB}[V/m] \approx 96 \sqrt{n_{pe} \left( {cm}^{-3} \right)}
	\end{equation}
	where $m_e$ is the electron mass and $\epsilon_0$ is the vacuum permittivity. It is easy to estimate the wave breaking field from equation (\ref{eq:wave_breaking_field2}). To generate a GV/m plasma wakefield to accelerate the witness beam, the plasma density $n_{pe}$ need to exceed $10^{14}$ ${cm}^{-3}$. At this density, the corresponding plasma wavelength $\lambda_{pe}$ is typically on the order of a few millimeters. According to \cite{lu2005limits}, the optimal wake would be obtained for $k_{pe} \sigma_z \approx \sqrt{2}$. This implies that the driver beam length $\sigma_z$ should be on the order of submillimeters (few hundreds of $\mu m$), to match the plasma wavelength and maximize wakefield excitation.
	
	Currently, the bunch lengths in existing heavy ion facilities are several meters long, making them inefficient for driving high amplitude plasma wakefields. To achieve effective wakefield excitation, two primary strategies have been proposed: compressing the entire driver beam or modulating it into microbunches for resonant excitation. However, since extensive research \cite{magneticcompression} has shown that beam compression poses significant challenges. Therefore, the following sections will focus on the latter approach.
	
\begin{figure}[h]
		\centering
		\begin{subfigure}[b]{0.48\textwidth}
			\centering
			\includegraphics[width=\linewidth]{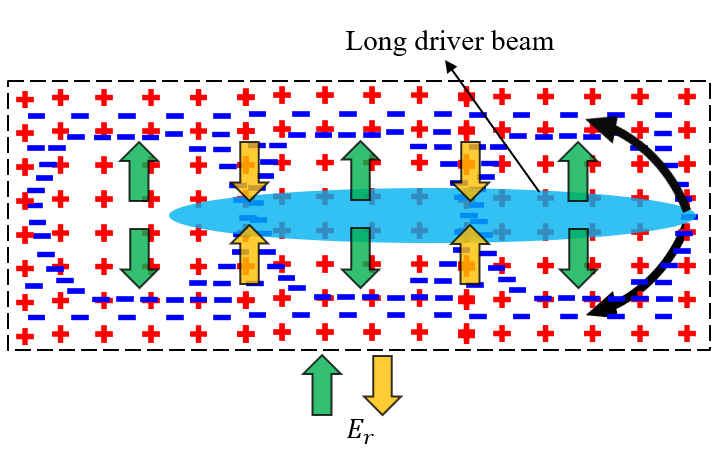}
			\caption{}
			\label{fig:Figure_2a}
		\end{subfigure}
		\hfill
		\begin{subfigure}[b]{0.48\textwidth}
			\centering
			\includegraphics[width=\linewidth]{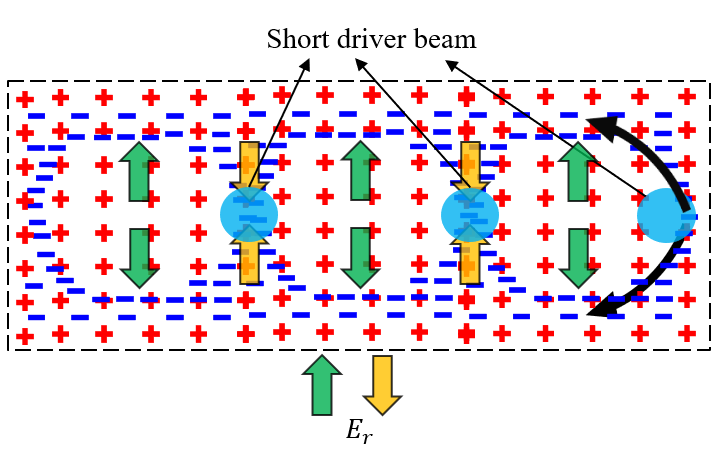}
			\caption{}
			\label{fig:Figure_2b}
		\end{subfigure}
		\caption{(Color) The formation of the wakefield excited by long driver beams. The head of long driver beam enters a plasma and excites a initial wakefield (a). Then, this wakefield acts on long driver beam itself and eventually generates a series of microbunches to resonantly drive wakefield (b).}
		\label{fig:Figure_2}
	\end{figure}

	Unlike these short electron beams or laser pulses can drive wakefields directly, the beam length of proton and heavy ion beams are too long to efficiently excite plasma wakefields. When a long heavy ion bunch ($\sigma_z \geq \lambda_{pe}$) enters a plasma, its head generates a wakefield that acts on its tail. This wakefield contains periodic focusing and defocusing regions, which modulate the beam density at the plasma wavelength ($ \lambda_{pe}$). These perturbations grow over time, leading to an unstable longitudinal modulation along the bunch. This process, known as the self-modulation instability (SMI) \cite{Kumar:2010bc,AWAKE:2023ssy,Caldwell:2011ir} eventually splits the long beam into a train of microbunches spaced by $ \lambda_{pe}/2$, thereby enabling resonant wakefield excitation, as illustrated in figure \ref{fig:Figure_2}. Unlike the electrostatic two-stream instability which arises from longitudinal relative motion between the beam and plasma, SMI is primarily driven by transverse wakefields acting back on the bunch itself.
	
	Several analytical models \cite{Schroeder:2011hkj} have been developed to describe the growth of self modulation instability (SMI) in plasma. However, the driver beam’s propagation in plasma is also affected by other instabilities, which may disrupt or even destroy the wakefield structure. By using seeded self modulation (SSM) \cite{AWAKE:2020stp,AWAKE:2022kmf,Lotov:2012ck,AWAKE:2017ulm}, these instabilities which are disadvantageous for the development of SMI can be mitigated. Rather than relying on random shot noise or beam imperfections, SMI can be triggered by employing a preceding bunch or a sharp leading-edge charge distribution to pre-generate transverse wakefields. This approach not only shortens the development distance for high amplitude wakefields but also suppresses competing instabilities.
	
	\section{Design method}
	
\begin{figure}[h]
		\centering
		\includegraphics[width=8.6cm]{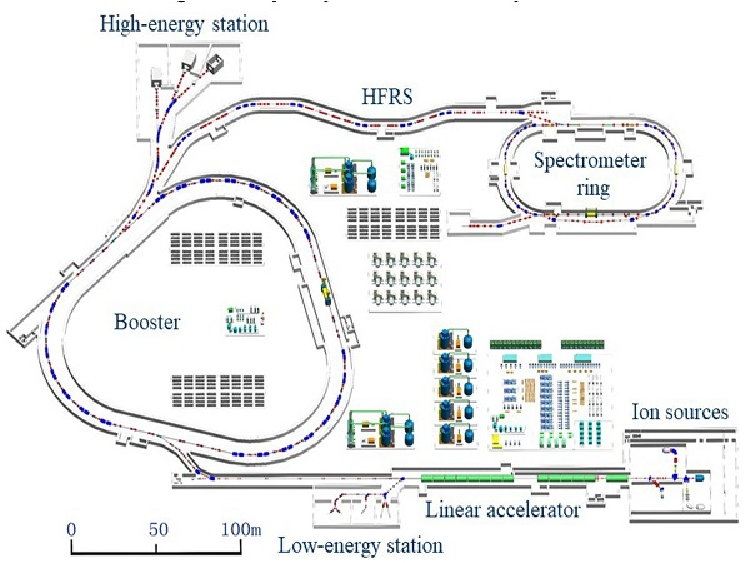}
		\caption{\label{fig:Figure_3}(Color) The layout of HIAF.}
	\end{figure}

	In this paper, we use the beam parameters from High Intensity heavy-ion Accelerator Facility (HIAF) in China \cite{Yang:2023hfx}. HIAF will be able to provide the highest pulse current of heavy ion beams in the world, with bunch kinetic energy reaching MJ level. Figure \ref{fig:Figure_3} shows the layout of HIAF. Scheduled for commissioning in 2025, HIAF is expected to provide high energy, high intensity heavy ion beams within a year or two. In the near future, it is an ideal platform for conducting heavy ion driven plasma wakefield acceleration experiments. The beam and plasma parameters are summarized in Table \ref{tab:HIAF_Bi_acc}.
	
	\begin{table}[h]
		
		\caption{\label{tab:HIAF_Bi_acc}The parameters of \(^{209}{Bi}^{83+}\) and witness beam and plasma for plasma acceleration in HIAF.}
		\begin{indented}
			\item[]\begin{tabular}{lc}
				\br
				Parameters &HIAF \\
				\mr
				Driver Beam& \(^{209}{Bi}^{83+}\)\\
				Energy(GeV/u)& 9.58 \\
				RMS Bunch Radius(mm)& 0.1\\
				Bunch Length(mm)& 0.314\\
				Relative energy spread(\%)& 0.035\\
				Plasma Density(${cm}^{-3}$)& $2.8 \times 10^{15}$\\
				Accelerating gradient(GV/m)& 6\\
				\mr
				Witness Beam& electron\\
				Energy(MeV)& 16 \\
				RMS Bunch Radius(mm)& 0.1\\
				Bunch Length(mm)& 0.314\\
				Relative energy spread(\%)& 0.035\\
				\mr
				Simulation grid step& 0.01\\
				Simulation window size(m)& 0.045\\
				particles in layer & 1000\\
				\br
			\end{tabular}
		\end{indented}
	\end{table}
	
	For a fully self-modulated driver beam in plasma wakefield acceleration, the dephasing length can be calculated based on the relative velocity difference between the driver and witness beam:
	
	\begin{equation}\label{eq:dephasing_length}
		L_d = \frac{\lambda_{pe} \beta_w}{4 \Delta \beta}
	\end{equation}
	where $\beta_d$ and $\beta_w$ are the velocities of the driver and witness beam, respectively, and $\Delta \beta$ is the relative velocity difference, which can be expressed as $\Delta \beta = \beta_w - \beta_d$. For the beam parameters in Table \ref{tab:HIAF_Bi_acc}, the dephasing length is estimated to be approximately 4.5 cm, which greatly limits the energy gain for witness electron beam in single acceleration stage. Electrons will exit the focusing and accelerating region, and enter the decelerating or defocusing area. This leads to a degradation in beam quality. A detailed discussion of these effects is provided in Section 4.
	
	To investigate optimal regimes for electron acceleration, plasma density variations are essential \cite{Petrenko:2015cxx,Braunmuller:2020aqw,AWAKE:2021vyl}. Plasma density variations can affect the quality of electron acceleration in three principal ways: (1) by modifying the dynamics of the heavy ion bunch and the growth rate of beam-plasma instabilities; (2) by affecting the trapping conditions for electrons; and (3) by influencing the structure and stability of the plasma wakefield. In this study, we assume that the self modulation instability of the Bismuth beam has fully developed that generate a train of microbunches and the witness electron bunch is already trapped. Under these assumptions, the main influence of the plasma density variations arises from the dephasing between the accelerating wakefield and the trapped electron bunch.
	
\begin{figure}[h]
		\centering
		\includegraphics[width=8.6cm]{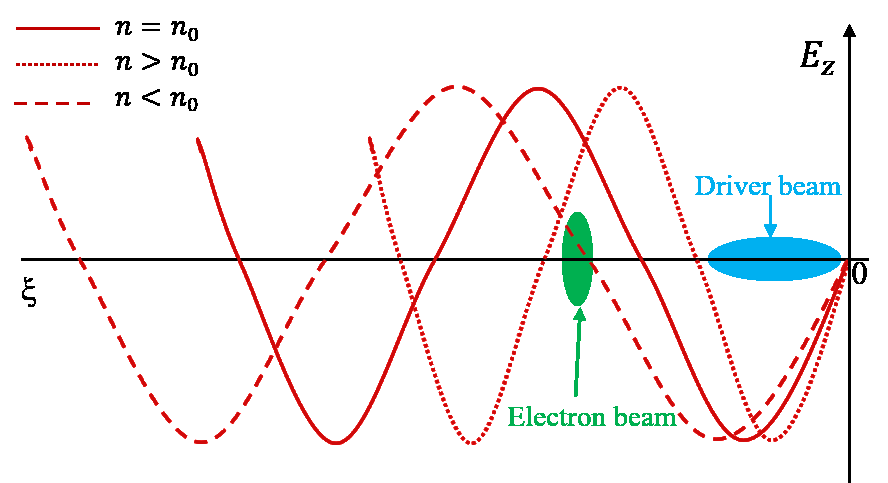}
		\caption{\label{fig:Figure_4}(Color) Phasing of the accelerated electron bunch in plasmas of the proper density (solid line), increased density (dotted line), and reduced density (dashed line).}
	\end{figure}

	During the development of SMI, the phase velocity of the plasma wave shifts backward as the instability evolves. In this stage, the wake is not yet suitable for efficient electron acceleration. Once the instability saturates and the microbunch structure is fully formed, the wakefield becomes appropriate for acceleration. At this stage, any variation in plasma density leads to forward or backward shifts in the phase of the plasma wave. This can be fetal to the witness beam, as illustrated in figure \ref{fig:Figure_4}.
	
	In a uniform plasma, electrons must reside in regions where both acceleration and transverse focusing region. When the plasma density increases, the plasma wavelength shortens, resulting in a forward phase shift of the local wakefield. This shift may move the electron bunch into a region with stronger accelerating fields or, undesirably, into a defocusing region, which can lead to beam loss. Conversely, a decrease in plasma density results in wavelength elongation, potentially placing the bunch in a region with weaker acceleration or even into a decelerating phase of the wave \cite{Lotov:2012ce,Katsouleas:1986zz}.
	
	A commonly method for plasma density variations is to linearly increase the plasma density, thereby continuously shortening the plasma wavelength. As the witness beam begins to overtake the driver beam, its phase is relocated so that it remains within the accelerating and focusing region. However, continuously increasing the plasma density leads to a mismatch between the RMS radius of the driver beam and the plasma wavenumber, violating the condition described in equation (\ref{eq:satisfy_condition2}). As a result, the wakefield strength excited by the driver beam in the plasma gradually weakens, eventually resulting in the disappearance of the accelerating field, which limits the energy gain of the witness beam. Therefore, simply increasing the plasma density cannot overcome the dephasing limitation.
	
\begin{figure}[h]
		\centering
		\includegraphics[width=8.6cm]{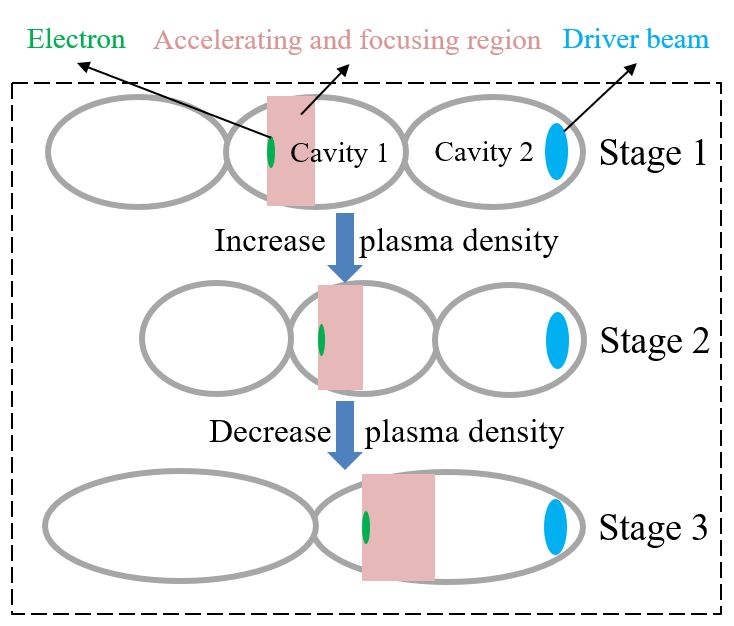}
		\caption{\label{fig:Figure_5}(Color) The principle of the alternating density gradients method.}
	\end{figure}

	In contrast, the method proposed in this paper uses alternating plasma density gradients to allow the witness beam to shift between adjacent accelerating cavities. This approach effectively mitigates the dephasing, enabling the witness beam to be accelerated continuously from the tail to the head of the driver beam. In this way, the entire wakefield generated by the driver beam in the plasma is fully utilized, achieving higher single stage energy gain, as illustrated in figure \ref{fig:Figure_5}.
	
	Initially, electrons are positioned near the beginning of the accelerating and focusing phase of cavity No.1. As the velocity of electrons exceeds that of the heavy ion driver, they gradually drift toward the decelerating phase. At this point, the plasma density is increased, thereby shortening the plasma wavelength and enabling that electrons are repositioned at the start of a new accelerating and focusing region. Once electrons again approaches the end of the accelerating phase, the plasma density is reduced to lengthen the plasma wavelength, allowing electrons to enter cavity No.2. By alternating the plasma density gradient, this process is repeated. Electrons will progressively move from the tail toward the head of the driver beam, ultimately achieving higher energy gain in one single stage.
	
	\begin{figure}[h]
		\centering
		\begin{subfigure}[b]{0.48\textwidth}
			\centering
			\includegraphics[width=\linewidth]{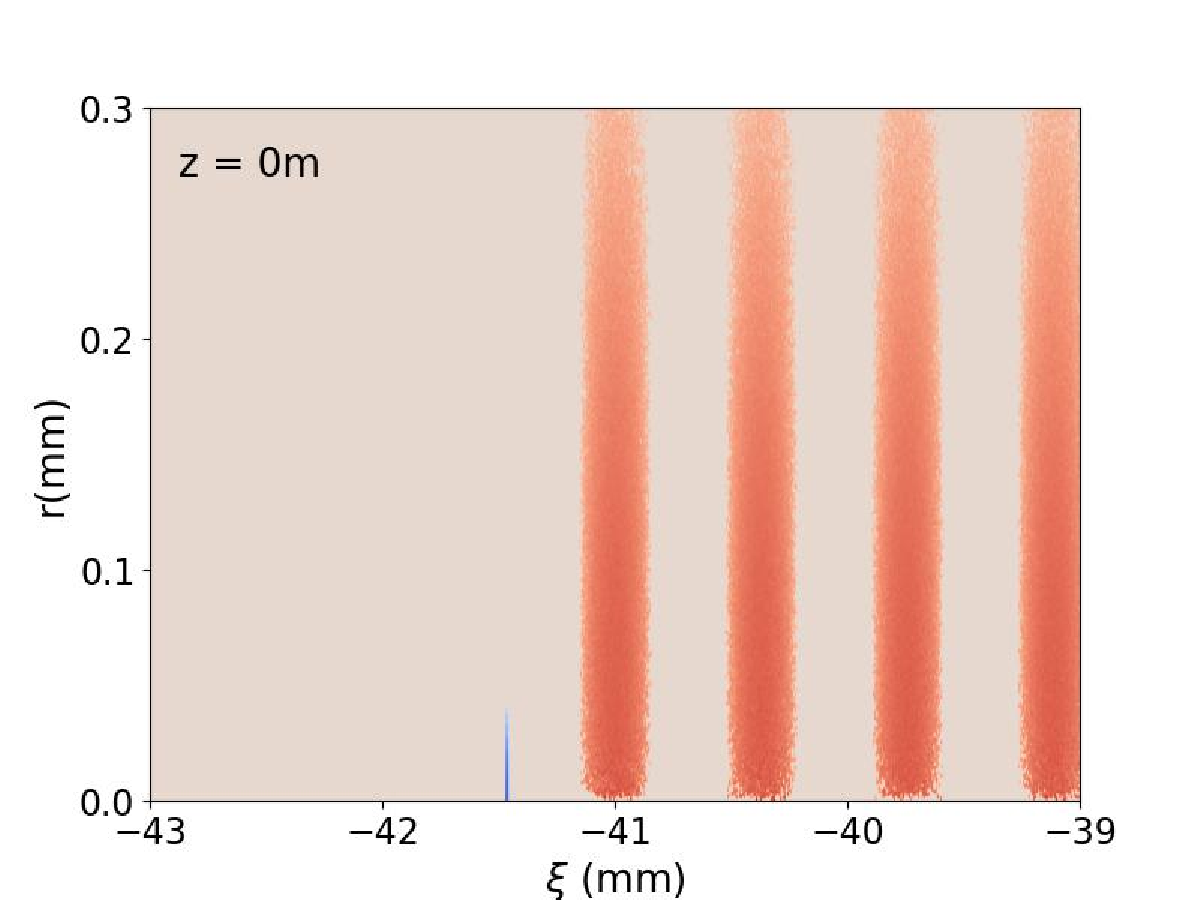}
			\caption{}
			\label{fig:Figure_6a}
		\end{subfigure}
		\hfill
		\begin{subfigure}[b]{0.48\textwidth}
			\centering
			\includegraphics[width=\linewidth]{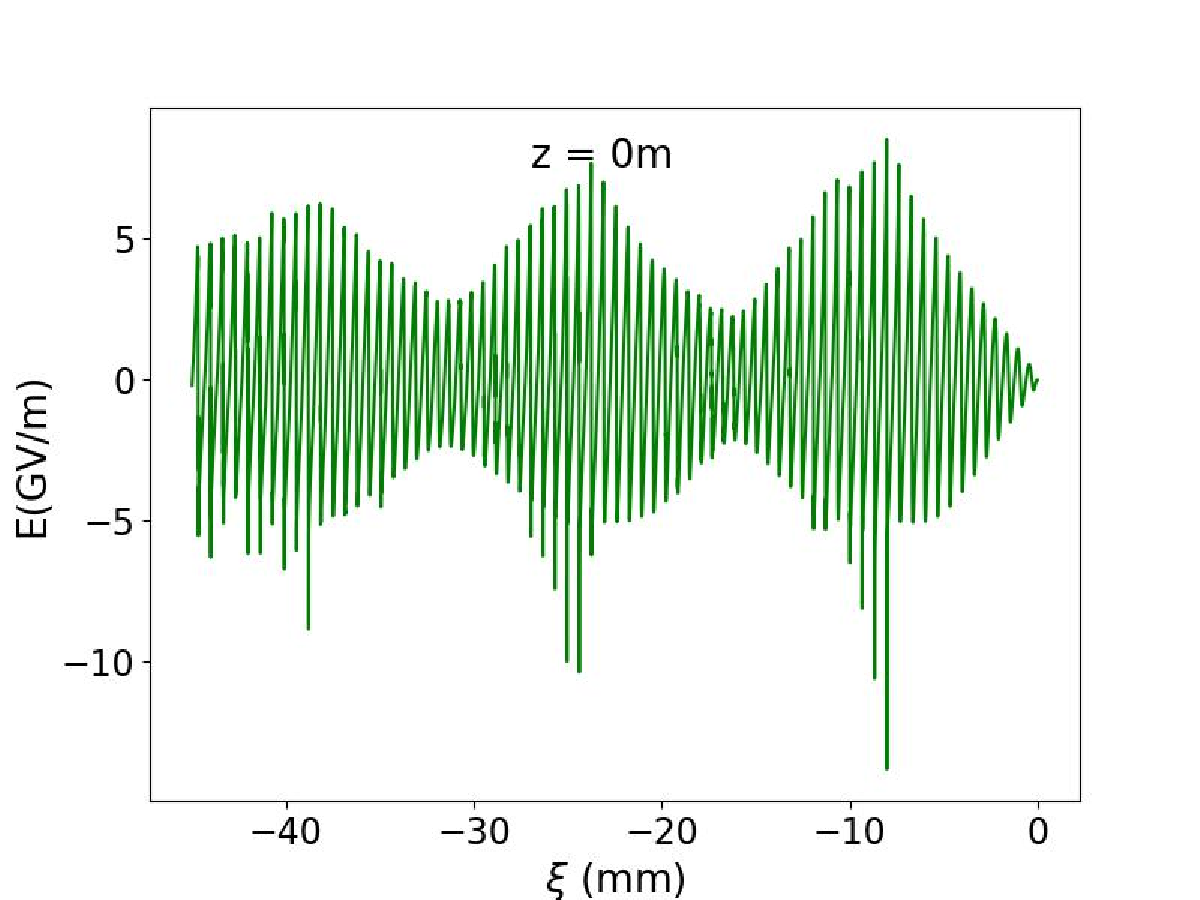}
			\caption{}
			\label{fig:Figure_6b}
		\end{subfigure}
		\vfill
		\begin{subfigure}[c]{0.48\textwidth}
			\centering
			\includegraphics[width=\linewidth]{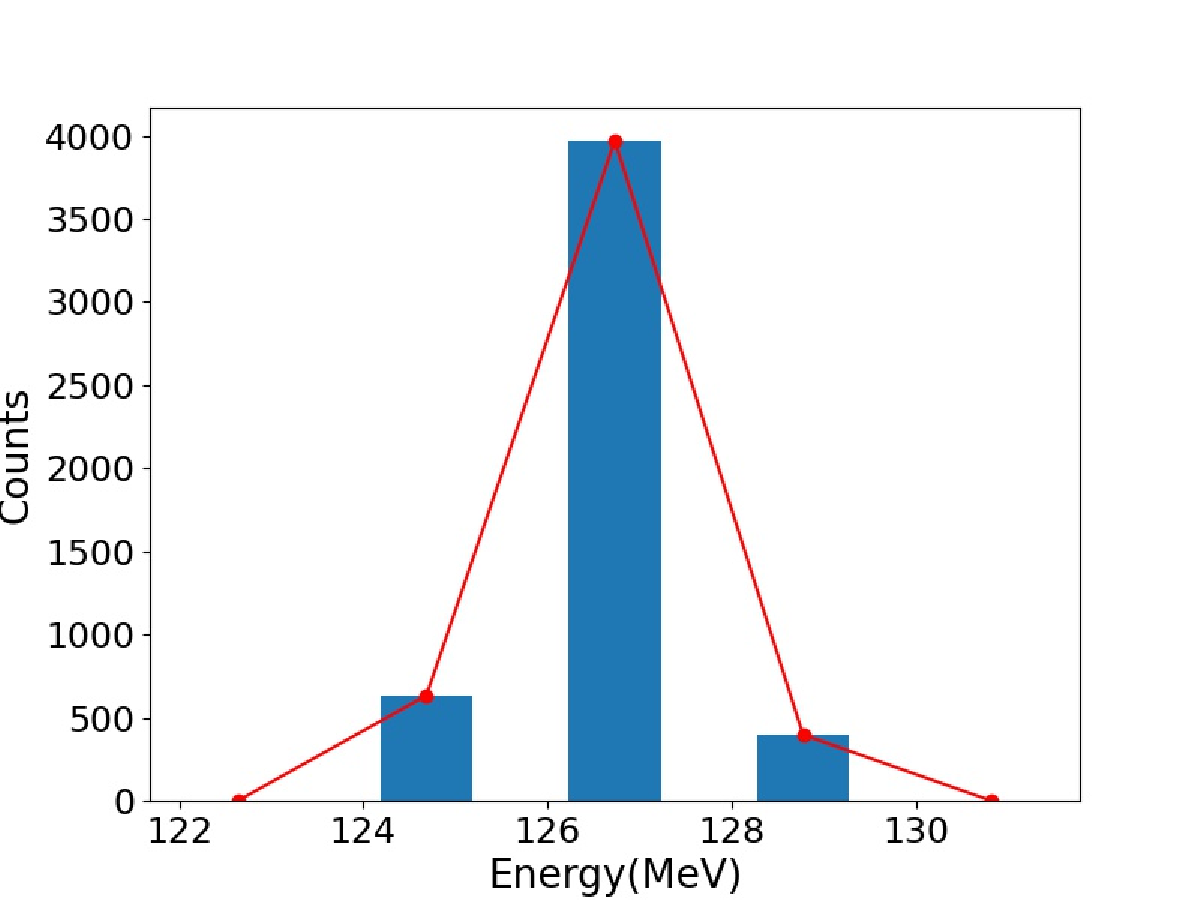}
			\caption{}
			\label{fig:Figure_6c}
		\end{subfigure}
		\caption{(Color) Without plasma density gradients, the initial electron (blue) and \(^{209}{Bi}^{83+}\) (red) beam distribution (a) , the initial wakefield amplitude in the beam head (b) for \(^{209}{Bi}^{83+}\) beam (RMS beam radius $=$ 0.1 mm) with co-moving coordinates $\xi$ and the electron energy distribution after propagating a distance of 6.6 cm (c).}
		\label{fig:Figure_6}
	\end{figure}
	
	\section{Simulations for electron acclecration}
	
	In this section, simulations of heavy ion driven plasma wakefield acceleration are conducted using the particle-in-cell (PIC) code LCODE \cite{Sosedkin:2016}. LCODE operates in two-dimensional (2D3V), supporting both planar and axisymmetric configurations. It employs a light-speed co-moving simulation window and utilizes the quasi-static approximation for plasma response calculations. Particle beams are modeled using fully relativistic macroparticles. The plasma can be treated either kinetically using macroparticles or via a fluid dynamics approach. The kinetic solver allows for the modeling of transversely inhomogeneous, hot, and non-neutral plasmas, as well as mobile ion dynamics. LCODE also has comprehensive diagnostics and graphical tools for visualization and analysis of simulation results.

	Using the parameters listed in Table \ref{tab:HIAF_Bi_acc}, wakefields greater than 5 GV / m can be achieved. without any plasma density gradients, electrons with an initial energy of 16 MeV can only be accelerated to 129 MeV within 6.6 cm, as shown in figure \ref{fig:Figure_6}.
	
	\begin{figure}[h]
		\centering
		\begin{subfigure}[b]{0.48\textwidth}
			\centering
			\includegraphics[width=\linewidth]{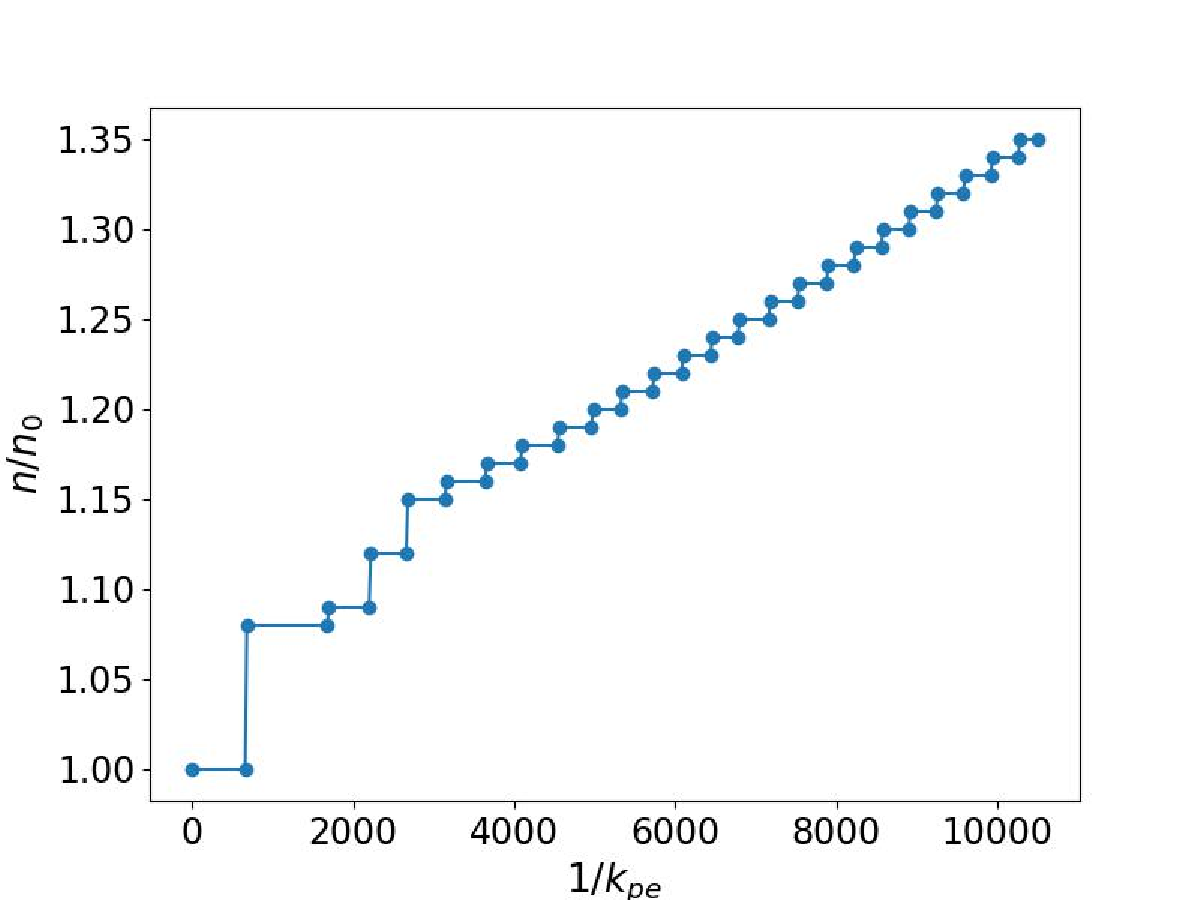}
			\caption{}
			\label{fig:Figure_7a}
		\end{subfigure}
		\hfill
		\begin{subfigure}[b]{0.48\textwidth}
			\centering
			\includegraphics[width=\linewidth]{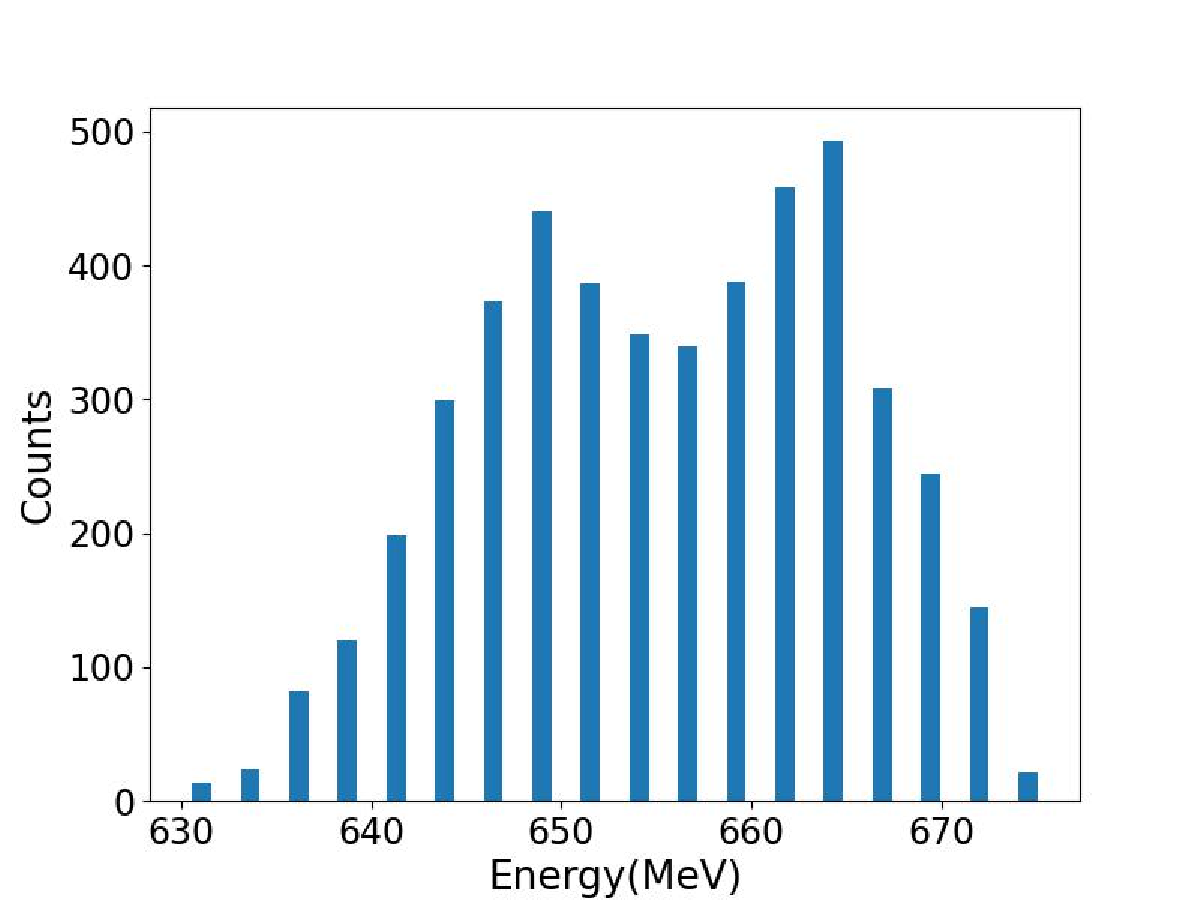}
			\caption{}
			\label{fig:Figure_7b}
		\end{subfigure}
		\vfill
		\begin{subfigure}[c]{0.48\textwidth}
			\centering
			\includegraphics[width=\linewidth]{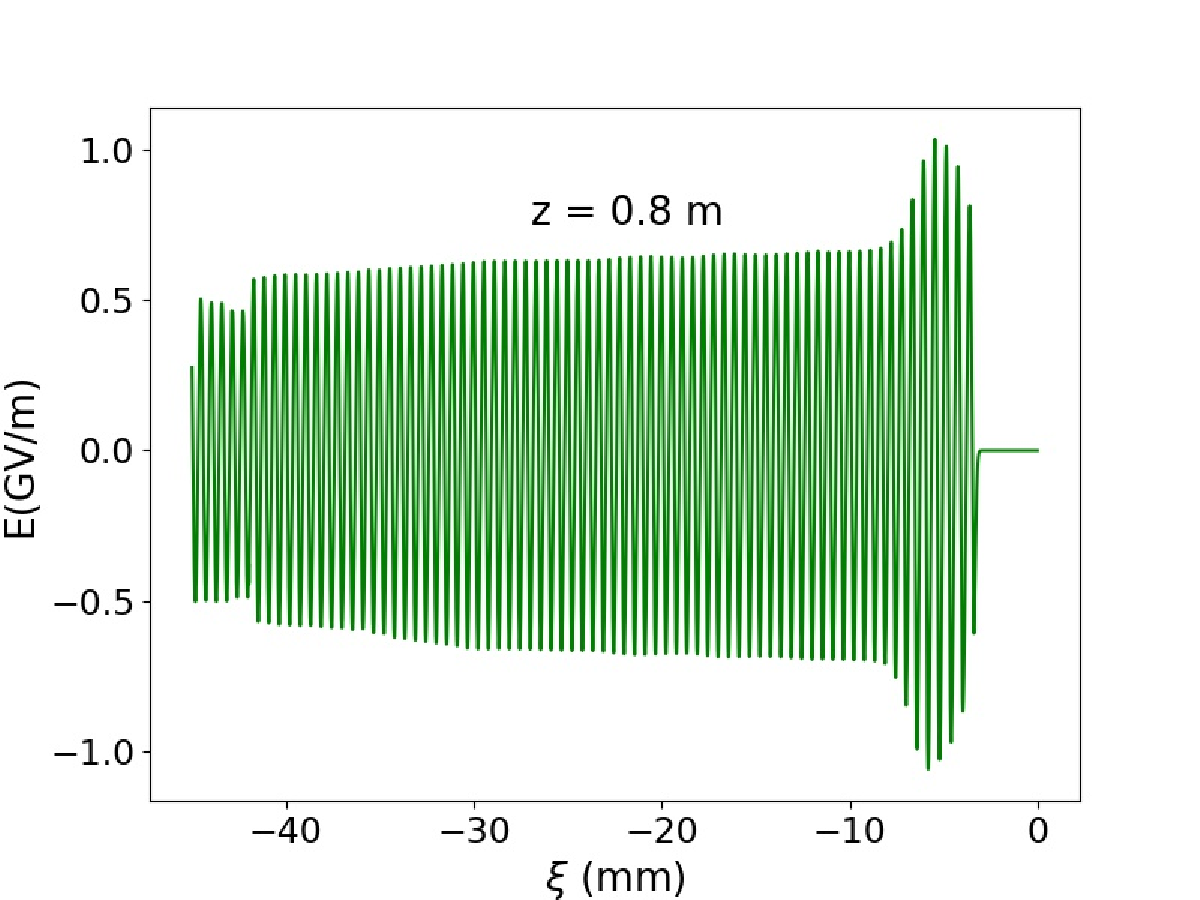}
			\caption{}
			\label{fig:Figure_7c}
		\end{subfigure}
		\caption{(Color) With the introduction of plasma density gradients, the plasma density variation (a), where $n_0 = 2.8 \times 10^{15} cm^{-3}$, the electron energy distribution after propagating a distance of 1 m (b) and the wakefield at the end of plasma (c).}
		\label{fig:Figure_7}
	\end{figure}
	
	In our previous work \cite{li2025numericalinvestigationsheavyion}, we adopted the method of continuously increasing the plasma density to mitigate dephasing and improve acceleration performance. Simulation results are presented in figure \ref{fig:Figure_7}. A narrow and short electron bunch can be accelerated from an initial energy of 16 MeV up to 675 MeV within a 1 m-long plasma, with a good energy spread of 1.5 \%. However, limited by the velocity of Bismush beam, the dephasing length remains short, resulting in an effective acceleration gradient of only 659 MV / m. Although we constantly increase the plasma density to prolong the dephasing length, it introduces a mismatch between the beam RMS radius and the plasma wavenumber. This mismatch leads to a gradual decay of the wakefield at the end of plasma, as shown in figure \ref{fig:Figure_7c}, where the wakefield strength at the end of the plasma is significantly weaker than at the entrance. Consequently, the energy gain of the electron beam is limited.
	
	\begin{figure}[h]
		\centering
		\begin{subfigure}[b]{0.48\textwidth}
			\centering
			\includegraphics[width=\linewidth]{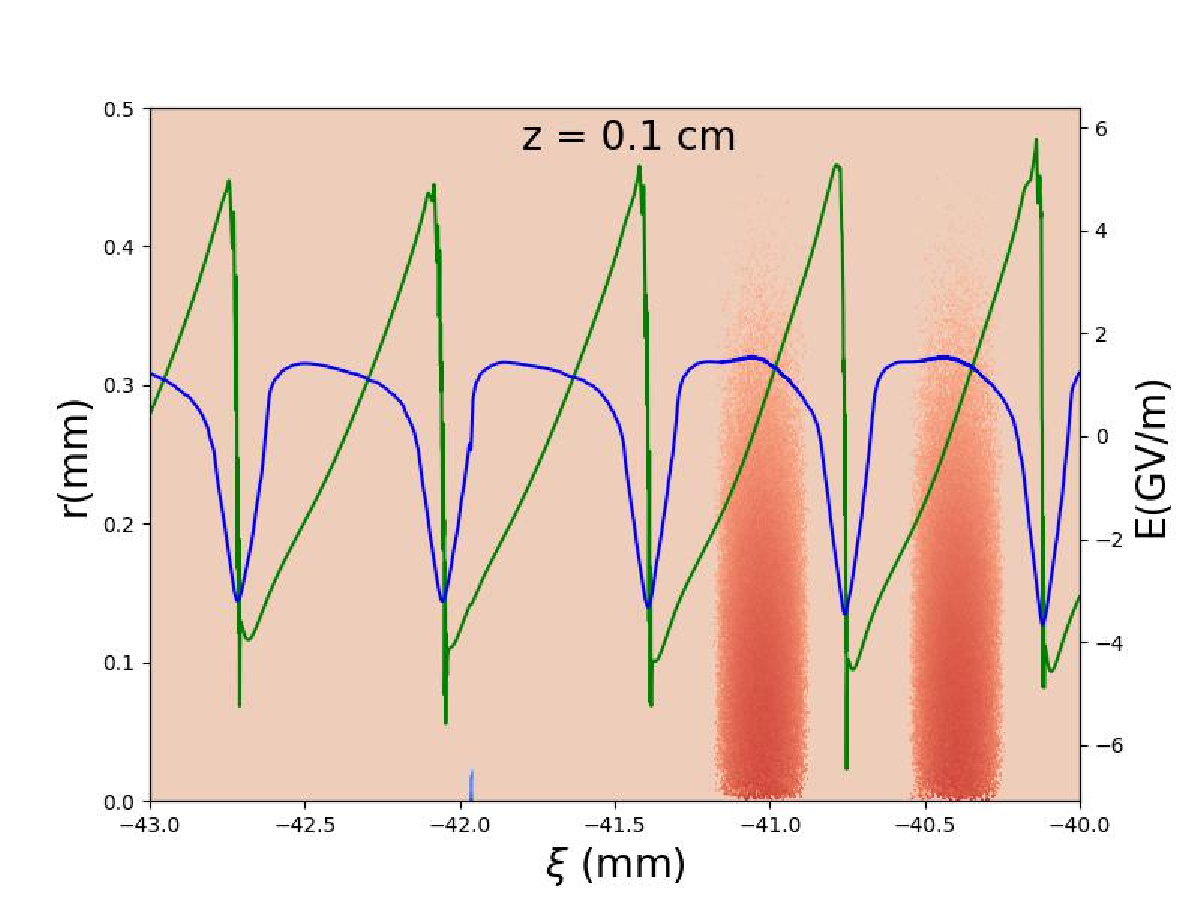}
			\caption{}
			\label{fig:Figure_8a}
		\end{subfigure}
		\hfill
		\begin{subfigure}[b]{0.48\textwidth}
			\centering
			\includegraphics[width=\linewidth]{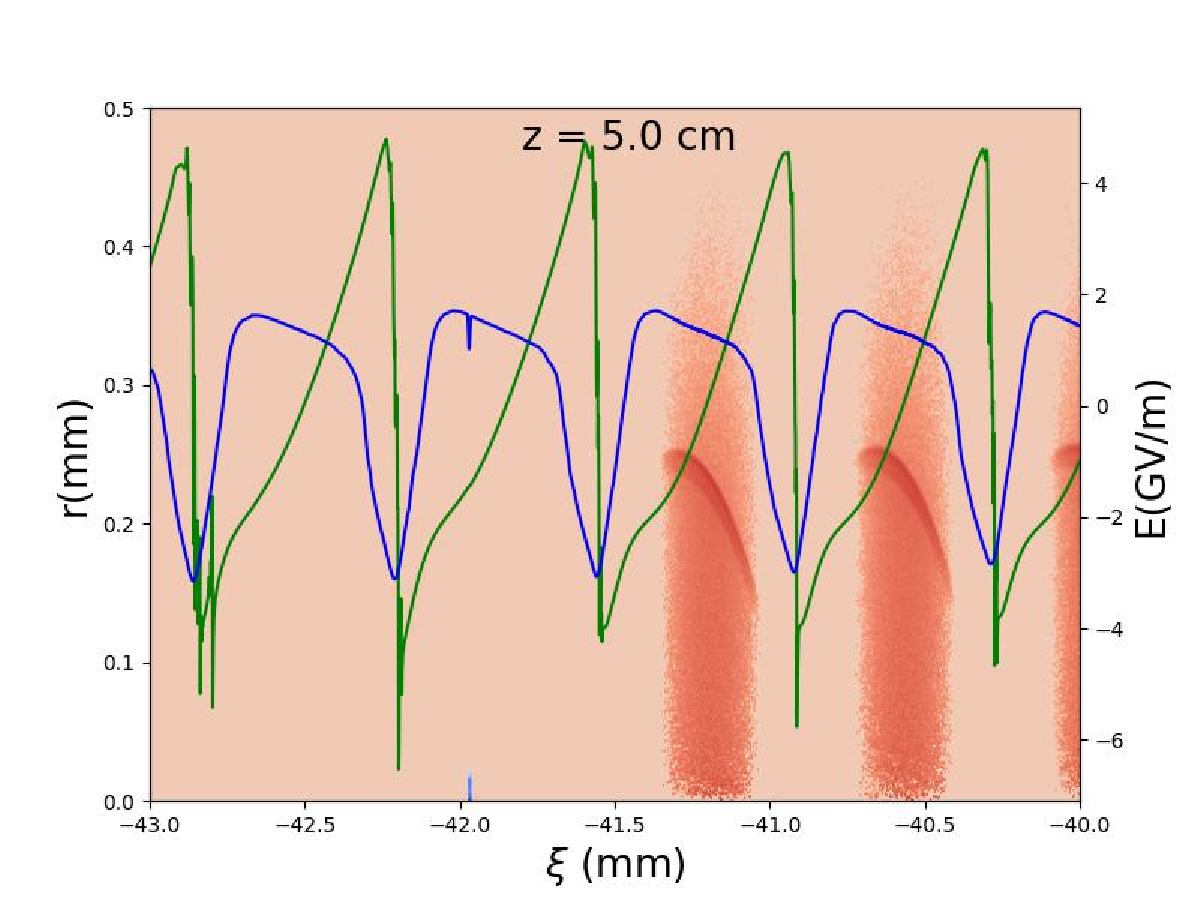}
			\caption{}
			\label{fig:Figure_8b}
		\end{subfigure}
		\vfill
		\begin{subfigure}[c]{0.48\textwidth}
			\centering
			\includegraphics[width=\linewidth]{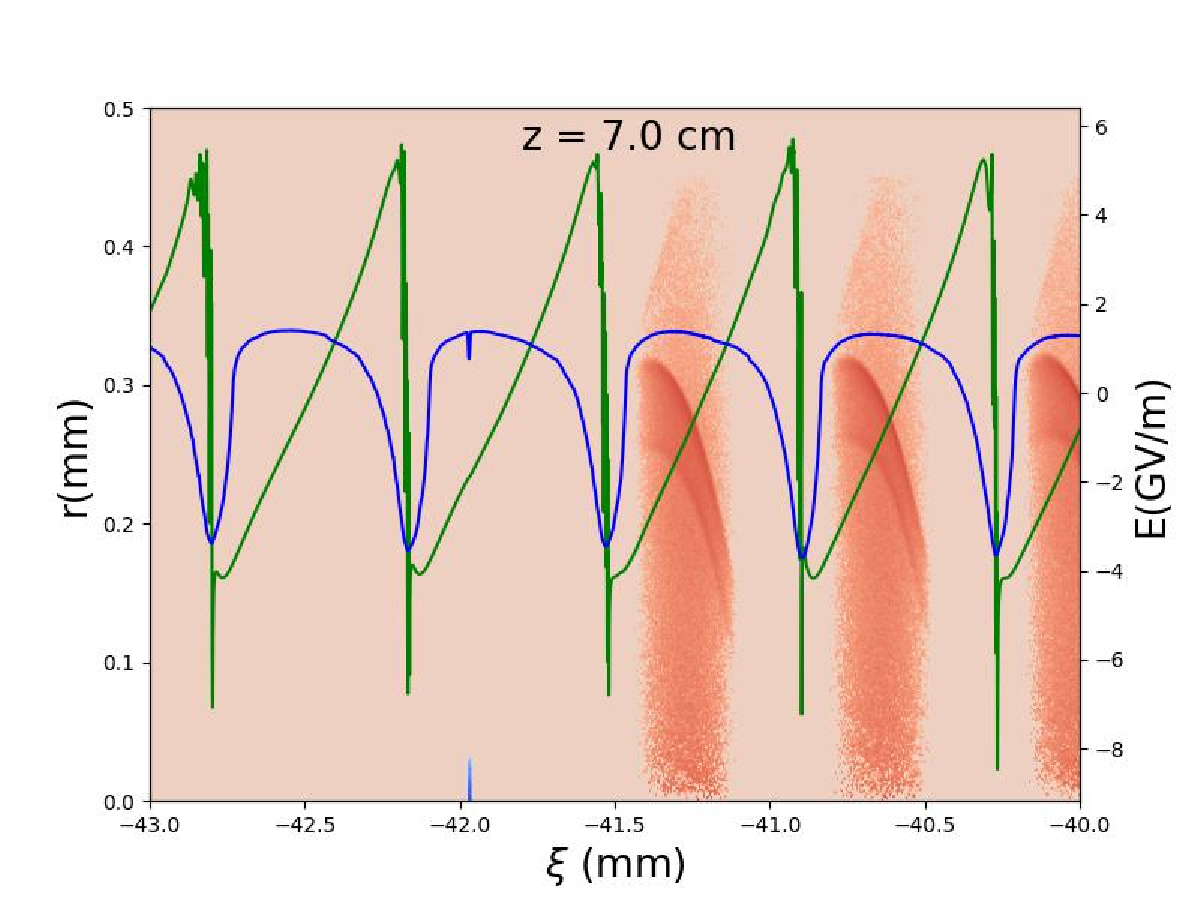}
			\caption{}
			\label{fig:Figure_8c}
		\end{subfigure}
		\caption{(Color) The initial electron (blue) and \(^{209}{Bi}^{83+}\) (red) beam distribution with co-moving coordinates $\xi$ in different time. Green line shows the longitudinal acceleration field and blue line shows the transverse field.}
		\label{fig:Figure_8}
	\end{figure}
	
\begin{figure}[h]
		\centering
		\begin{subfigure}[b]{0.48\textwidth}
			\centering
			\includegraphics[width=\linewidth]{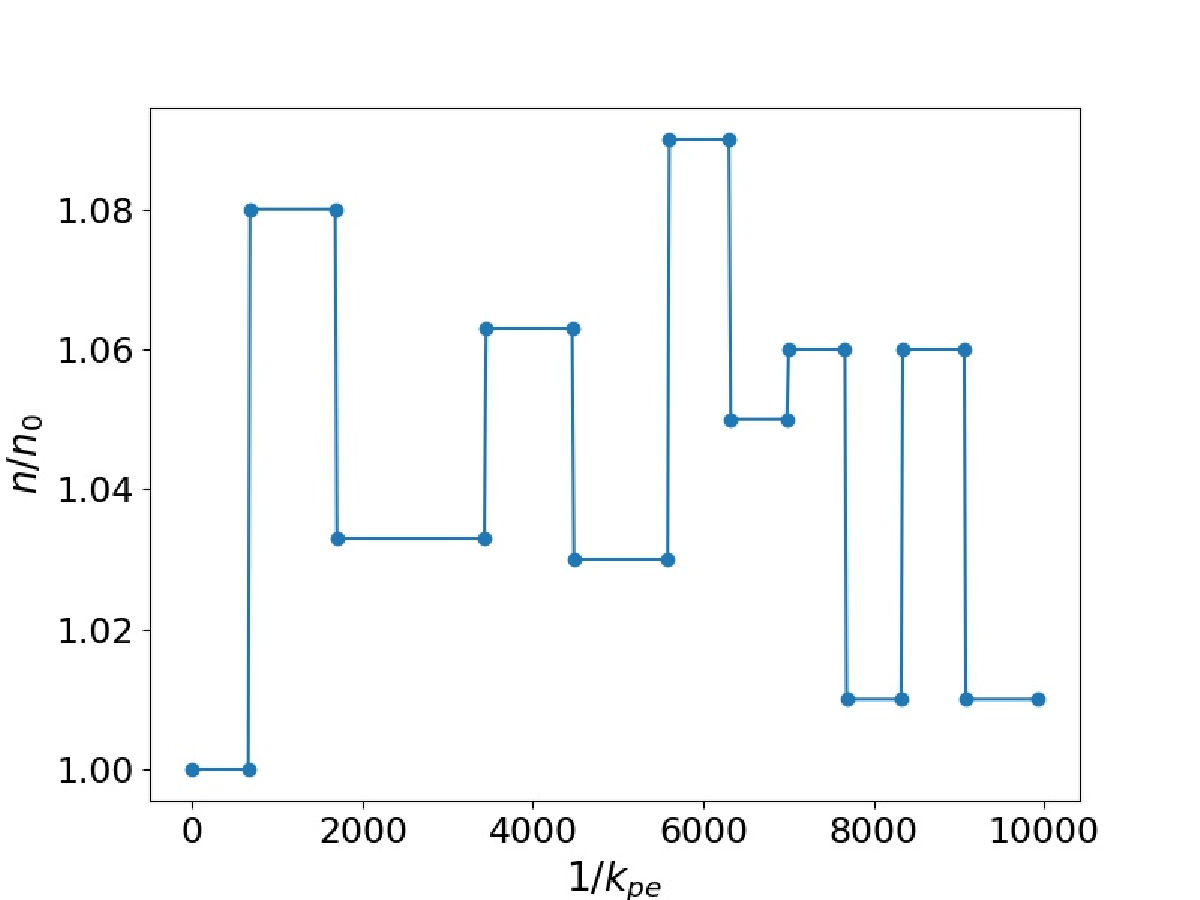}
			\caption{}
			\label{fig:Figure_9a}
		\end{subfigure}
		\hfill
		\begin{subfigure}[b]{0.48\textwidth}
			\centering
			\includegraphics[width=\linewidth]{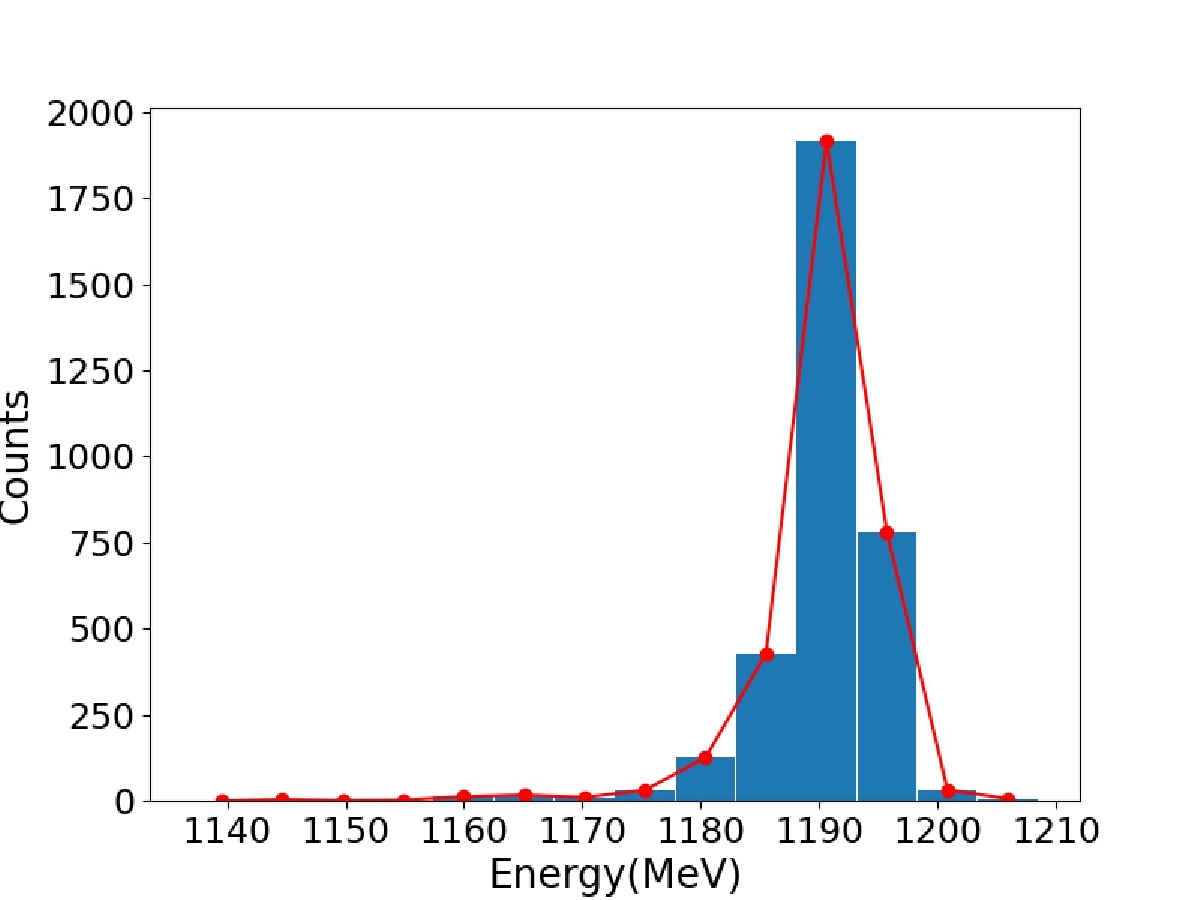}
			\caption{}
			\label{fig:Figure_9b}
		\end{subfigure}
		\caption{(Color) With the introduction of plasma density gradients, the plasma density variation (a), where $n_0 = 2.8 \times 10^{15} cm^{-3}$, the electron energy distribution after propagating a distance of 1 m (b).}
		\label{fig:Figure_9}
	\end{figure}

	Using the same beam and initial plasma parameters, we apply the alternating density gradients method to optimize the acceleration process. Figure \ref{fig:Figure_8} shows the details in our simulation. Initially, electrons are located at the beginning of the accelerating and focusing region, shown in figure \ref{fig:Figure_8a}. Electrons are easy to catch up with the Bismush bunch and reach the decelerating region, shown in figure \ref{fig:Figure_8b}. At this point, the plasma density is increased, making the electrons to enter a subsequent accelerating and focusing phase, shown in figure \ref{fig:Figure_8c}. Then, as the electrons are about to dephasing again, the plasma density is reduced, shifting them into a preceding accelerating phase. This process will be repeated until the electrons traverse from the tail to the head of the heavy ion beam, fully utilizing the plasma wakefield excited by the heavy ion beam. Simulation results show that, under a properly designed alternating density gradient profile, electrons can be accelerated up to 1.2 GeV in a distance of 0.99 m, resulting in an energy spread of 1.2 \%, shown in figure \ref{fig:Figure_9}.
	
\begin{figure}[h]
		\centering
		\includegraphics[width=8.6cm]{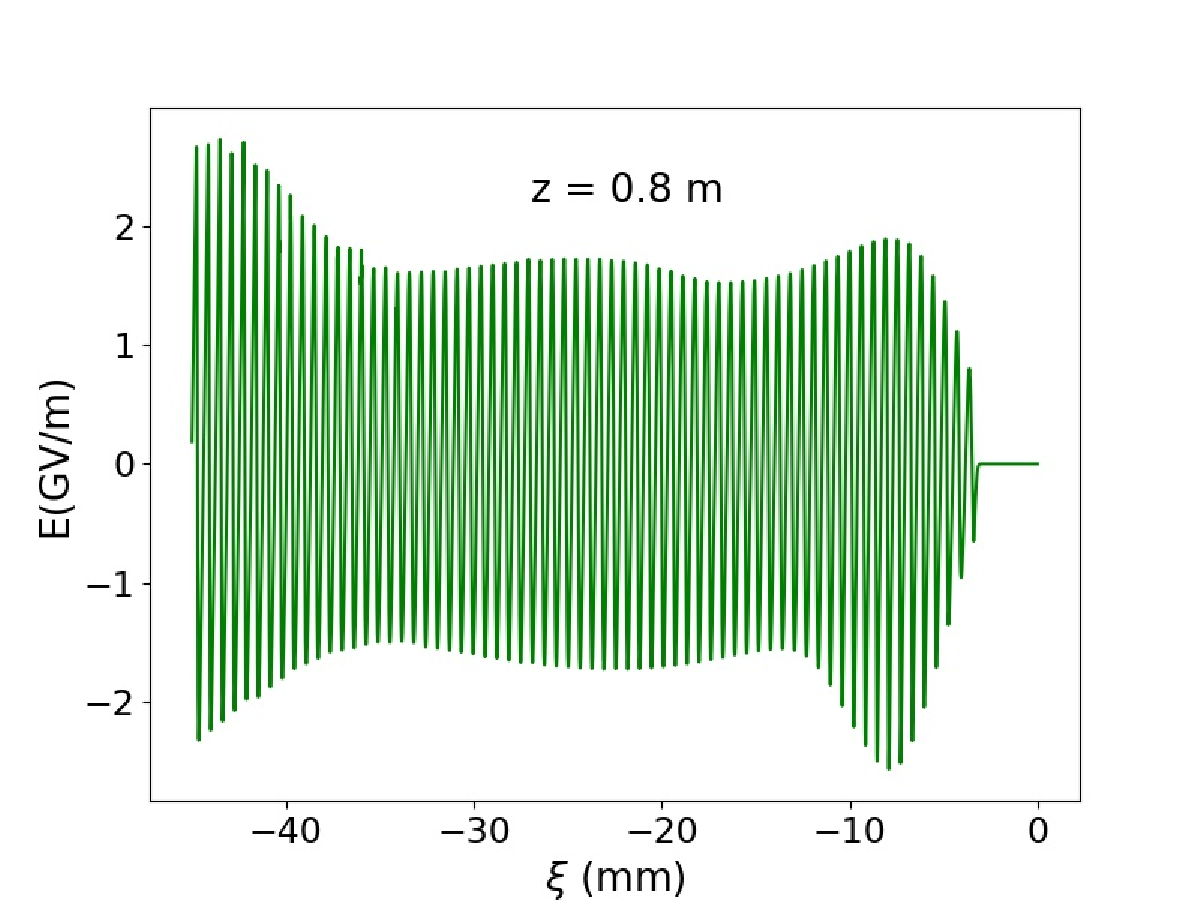}
		\caption{\label{fig:Figure_10}(Color) The wakefield at the end of plasma.}
	\end{figure}

	Due to limitations in computing resources, the simulation time was insufficient to allow the electrons to fully traverse the entire heavy ion beam. Nevertheless, we will continue to optimize the beam and plasma parameters in future studies. Compared to our previous work, the effective acceleration gradient using the alternating plasma density gradient method is approximately twice that of the linearly increasing plasma density. Furthermore, the presence of a strong wakefield near the plasma exit, as shown in \ref{fig:Figure_10}, indicates that the electron bunch can continue to gain energy beyond the simulated range. This confirms the potential of this method to achieve even higher energy gains in extended plasma lengths.
	
	\section{Conclusions}
	
	In summary, heavy ion beam possess some potential benefits in plasma wakefield acceleration, particularly due to their high kinetic energy and ability to generate strong wakefields. However, their relatively low velocity leads to a short dephasing length, which severely limits the maximum energy gain that the witness beam can achieve. The traditional method of linearly increasing plasma density can modestly extendg the dephasing length and increase the energy gain of the witness beam. However, this method introduces a gradual mismatch between the RMS raidus of driver beam and the plasma wavenumber, causing the wakefield strength to Weaken and eventually vanish. Therefore, the full potential of heavy ion drivers cannot be realized using this strategy.
	
	In this paper, we propose an alternative approach based on alternating density gradients, electron will switch between adjacent acceleration cavities and keep being accelerated, which may effectively extends the dephasing length and significantly increase the energy of the witness beam. Simulation results show that with a properly designed alternating density gradients modulation, electrons can be accelerated up to 1.14 GeV in a distance of 0.92 m, resulting in an energy spread of 1.3 \%. The efficient acceleration gradient is approximately twice that using the traditional method. Moreover, the strong wakefield observed at the plasma exit indicates that further acceleration is possible beyond the simulated domain, suggesting the potential for even higher energies with extended plasma stages.
	
	\section{Acknowledgments}
	
	The authors would like to appreciate the revision comments from Y.J.Yuan. This work is supported by the China National Funds for Distinguished Young Scientists (Grant No. 12425501).
	
	\section*{References}
	\providecommand{\newblock}{}


\begin{thebibliography}{10}
		\expandafter\ifx\csname url\endcsname\relax
		\def\url#1{{\tt #1}}\fi
		\expandafter\ifx\csname urlprefix\endcsname\relax\def\urlprefix{URL }\fi
		\providecommand{\eprint}[2][]{\url{#2}}
		
		\bibitem{FCC:2018byv}
		Abada A {\em et~al.\/} (FCC) 2019 {\em Eur. Phys. J. C\/} {\bf 79} 474
		
		\bibitem{Tajima:1979bn}
		Tajima T and Dawson J~M 1979 {\em Phys. Rev. Lett.\/} {\bf 43} 267--270
		
		\bibitem{Esarey:2009zz}
		Esarey E, Schroeder C~B and Leemans W~P 2009 {\em Rev. Mod. Phys.\/} {\bf 81}
		1229--1285
		
		\bibitem{Pukhov:2002otp}
		Pukhov A and Meyer-ter Vehn J 2002 {\em Appl. Phys. B\/} {\bf 74} 355--361
		
		\bibitem{Lu:2006nz}
		Lu W, Tzoufras M, Joshi C, Tsung F~S, Mori W~B, Vieria J, Fonseca R~A and Silva
		L~O 2007 {\em Phys. Rev. ST Accel. Beams\/} {\bf 10} 061301
		(\textit{Preprint} \eprint{physics/0612227})
		
		\bibitem{Picksley:2024cdd}
		Picksley A {\em et~al.\/} 2024 {\em Phys. Rev. Lett.\/} {\bf 133} 255001
		(\textit{Preprint} \eprint{2408.00740})
		
		\bibitem{Chen:1984up}
		Chen P, Dawson J~M, Huff R~W and Katsouleas T~C 1985 {\em Phys. Rev. Lett.\/}
		{\bf 54} 693--696 [Erratum: Phys.Rev.Lett. 55, 1537 (1985)]
		
		\bibitem{Blue:2003nk}
		Blue B~E {\em et~al.\/} 2003 {\em Phys. Rev. Lett.\/} {\bf 90} 214801
		
		\bibitem{Blumenfeld:2007ph}
		Blumenfeld I {\em et~al.\/} 2007 {\em Nature\/} {\bf 445} 741--744
		
		\bibitem{Litos:2014yqa}
		Litos M {\em et~al.\/} 2014 {\em Nature\/} {\bf 515} 92--95
		
		\bibitem{Ruth:1984pz}
		Ruth R~D, Chao A~W, Morton P~L and Wilson P~B 1985 {\em Part. Accel.\/} {\bf
			17} 171
		
		\bibitem{Muggli:2016pou}
		Muggli P and Bracco C (AWAKE) 2016 {AWAKE, the Advanced Proton Driven Plasma
			Wakefield Acceleration Experiment} {\em {7th International Particle
				Accelerator Conference}\/} p WEPMY019
		
		\bibitem{Adli:2016rwp}
		Adli E and Muggli P 2016 {\em Rev. Accel. Sci. Tech.\/} {\bf 09} 85--104
		
		\bibitem{Caldwell:2015rkk}
		Caldwell A {\em et~al.\/} 2016 {\em Nucl. Instrum. Meth. A\/} {\bf 829} 3--16
		(\textit{Preprint} \eprint{1511.09032})
		
		\bibitem{AWAKE:2018gdq}
		Adli E {\em et~al.\/} (AWAKE) 2018 {\em Nature\/} {\bf 561} 363--367
		(\textit{Preprint} \eprint{1808.09759})
		
		\bibitem{li2025numericalinvestigationsheavyion}
		Li J, Yang J, Xia G, Liu J, Zhan W and Zhu R 2025 Numerical investigations of
		heavy ion driven plasma wakefield acceleration (\textit{Preprint}
		\eprint{2506.14132}) \urlprefix\url{https://arxiv.org/abs/2506.14132}
		
		\bibitem{AWAKE:2023pqh}
		Verra L {\em et~al.\/} (AWAKE) 2024 {\em Phys. Rev. E\/} {\bf 109} 055203
		(\textit{Preprint} \eprint{2312.13883})
		
		\bibitem{lu2005limits}
		Lu W, Huang C, Zhou M, Mori W and Katsouleas T 2005 {\em Physics of plasmas\/}
		{\bf 12}
		
		\bibitem{magneticcompression}
		Liu C, Xia G~X, Zhuang J~J, Lu X~Y, Zhang B~C and Zhao K 2006 {\em Qiangjiguang
			Yu Lizishu/High Power Laser and Particle Beams\/} {\bf 18} 139--142
		
		\bibitem{Kumar:2010bc}
		Kumar N, Pukhov A and Lotov K 2010 {\em Phys. Rev. Lett.\/} {\bf 104} 255003
		(\textit{Preprint} \eprint{1003.5816})
		
		\bibitem{AWAKE:2023ssy}
		Verra L {\em et~al.\/} (AWAKE) 2023 {\em Phys. Plasmas\/} {\bf 30} 083104
		(\textit{Preprint} \eprint{2305.05478})
		
		\bibitem{Caldwell:2011ir}
		Caldwell A and Lotov K 2011 {\em Phys. Plasmas\/} {\bf 18} 103101
		(\textit{Preprint} \eprint{1105.1292})
		
		\bibitem{Schroeder:2011hkj}
		Schroeder C~B, Benedetti C, Esarey E, Gr\"uner F~J and Leemans W~P 2011 {\em
			Phys. Rev. Lett.\/} {\bf 107} 145002 (\textit{Preprint} \eprint{1108.1564})
		
		\bibitem{AWAKE:2020stp}
		Batsch F {\em et~al.\/} (AWAKE) 2021 {\em Phys. Rev. Lett.\/} {\bf 126} 164802
		(\textit{Preprint} \eprint{2012.09676})
		
		\bibitem{AWAKE:2022kmf}
		Verra L {\em et~al.\/} (AWAKE) 2022 {\em Phys. Rev. Lett.\/} {\bf 129} 024802
		(\textit{Preprint} \eprint{2203.13752})
		
		\bibitem{Lotov:2012ck}
		Lotov K~V, Lotova G~Z, Lotov V~I, Upadhyay A, T\"uckmantel T, Pukhov A and
		Caldwell A 2013 {\em Phys. Rev. ST Accel. Beams\/} {\bf 16} 041301
		(\textit{Preprint} \eprint{1204.3444})
		
		\bibitem{AWAKE:2017ulm}
		Muggli P {\em et~al.\/} (AWAKE) 2017 {\em Plasma Phys. Control. Fusion\/} {\bf
			60} 014046 (\textit{Preprint} \eprint{1708.01087})
		
		\bibitem{Yang:2023hfx}
		Yang J, Sun L and Yuan Y 2023 {\em JACoW\/} {\bf CYCLOTRONS2022} MOAI01
		
		\bibitem{Petrenko:2015cxx}
		Petrenko A, Lotov K and Sosedkin A 2016 {\em Nucl. Instrum. Meth. A\/} {\bf
			829} 63--66 (\textit{Preprint} \eprint{1511.04360})
		
		\bibitem{Braunmuller:2020aqw}
		Braunm\"uller F, Nechaeva T and Collboration A (AWAKE) 2020 {\em Phys. Rev.
			Lett.\/} {\bf 125} 264801 (\textit{Preprint} \eprint{2007.14894})
		
		\bibitem{AWAKE:2021vyl}
		Morales~Guzm\'an P~I {\em et~al.\/} (AWAKE) 2021 {\em Phys. Rev. Accel.
			Beams\/} {\bf 24} 101301 (\textit{Preprint} \eprint{2107.11369})
		
		\bibitem{Lotov:2012ce}
		Lotov K~V, Pukhov A and Caldwell A 2013 {\em Phys. Plasmas\/} {\bf 20} 013102
		(\textit{Preprint} \eprint{1205.3388})
		
		\bibitem{Katsouleas:1986zz}
		Katsouleas T~C 1986 {\em Phys. Rev. A\/} {\bf 33} 2056--2064
		
		\bibitem{Sosedkin:2016}
		Sosedkin A~P and Lotov K~V 2016 {\em Nucl. Instrum. Meth. A\/} {\bf 829}
		350--352 (\textit{Preprint} \eprint{1511.04193})
		
	\end{thebibliography}
\end{document}